\documentclass[aps,prb,twocolumn,showpacs,preprintnumbers,amsmath,amssymb,groupedaddress,noshowkeys]{revtex4}
\usepackage{txfonts}
\usepackage[dvipdfm]{graphicx}
\usepackage{bm}
\def\rnum#1{\expandafter{\romannumeral #1}} 
\def\Rnum#1{\uppercase\expandafter{\romannumeral #1}}

\newfont{\bg}{cmr10 scaled\magstep4}

\newcommand{\bigzerou}{\smash{\lower1.8ex\hbox{\bg 0}}}

\begin{document}

\title{Generation of spin-polarized current using \\
multi-terminated quantum dot with spin-orbit interaction}
\author{Tomohiro Yokoyama}
\email[E-mail me at: ]{tyokoyam@rk.phys.keio.ac.jp}

\author{Mikio Eto}
\affiliation{Faculty of Science and Technology, Keio University,
3-14-1 Hiyoshi, Kohoku-ku, Yokohama 223-8522, Japan}
\date{\today}

\begin{abstract}
We theoretically examine generation of spin-polarized current
using multi-terminated quantum dot with spin-orbit interaction.
First, a two-level quantum dot is analyzed as a minimal model,
which is connected to $N$ ($\ge 2$) external leads via tunnel
barriers. When an unpolarized current is injected to the quantum dot
from a lead, a polarized current is ejected to others, similarly
to the spin Hall effect. In the absence of magnetic field,
the generation of spin-polarized current requires $N \ge 3$.
The polarization is markedly enhanced by resonant tunneling when
the level spacing in the quantum dot is smaller than the level
broadening due to the tunnel coupling to the leads.
In a weak magnetic field, the orbital magnetization
creates a spin-polarized current even in the two-terminal geometry
($N=2$). The numerical study for generalized situations confirms
our analytical result using the two-level model.
\end{abstract}
\maketitle

\section{INTRODUCTION}
\label{sec:Intro}

The spin-orbit (SO)
interaction in semiconductors has been
studied extensively from viewpoints of its fundamental research
and application to spin-based electronics,
``spintronics.''~\cite{Zutic}
For conduction electrons in direct-gap semiconductors,
an external potential $U({\bm r})$ results in the
Rashba SO interaction~\cite{Rashba,Rashba2}
\begin{equation}
H_{\rm RSO} =
\frac{\lambda}{\hbar} {\bm \sigma} \cdot
\left[{\bm p} \times {\bm \nabla} U({\bm r}) \right],
\label{eq:SORashba}
\end{equation}
where ${\bm p}$ is the momentum operator and
${\bm \sigma}$ is the Pauli matrices indicating
the electron spin ${\bm s}={\bm \sigma}/2$.
The coupling constant $\lambda$ is markedly
enhanced by the band effect, particularly
in narrow-gap semiconductors, such as
InAs and InSb.~\cite{Winkler,Nitta}
The bulk inversion symmetry is broken in
compound semiconductors, which gives rise to another
type of SO interaction, the Dresselhaus SO
interaction.~\cite{Dresselhaus} It is given by
\begin{eqnarray}
H_{\rm DSO}=\frac{\lambda'}{\hbar}
\bigl[ p_x(p_y^2-p_z^2)\sigma_x+
p_y(p_z^2-p_x^2)\sigma_y
\nonumber \\
+p_z(p_x^2-p_y^2)\sigma_z \bigr].
\label{eq:Dresselhaus}
\end{eqnarray}

In the presence of SO interaction, the spin Hall effect (SHE)
is one of the most important phenomena for the application
to the spintronics. It produces a spin current traverse to
an electric field applied by the bias voltage.
There are two types of SHE. One is an intrinsic SHE,
which is induced by the drift motion of carriers in the
SO-split band structures. It creates a dissipationless spin
current.~\cite{Murakami,Wunderlich,Sinova}
The other is an extrinsic SHE caused by the
spin-dependent scattering of electrons by
impurities.~\cite{Dyakonov}
Kato {\it et al}.\ observed the spin accumulation at
sample edges traverse to the current,~\cite{Kato}
which is ascribable to the extrinsic SHE with
$U({\bm r})$ being the screened Coulomb potential by
charged impurities in Eq.\ (\ref{eq:SORashba}).~\cite{Engel}
The extrinsic SHE is usually understood semi-classically
in terms of skew scattering and side-jump effect.

In our previous studies,~\cite{EYjpsj1,YEprb}
we theoretically examined the extrinsic SHE in semiconductor
heterostructures due to the scattering by single artificial
potential. The potential created by antidots, STM tips, and
others, is electrically tunable. We adopted the quantum
mechanical scattering theory for this problem.
When the potential is axially symmetric in two dimensions,
$U(r)$ with $r=\sqrt{x^2+y^2}$ in the $xy$ plane, electrons
feel the potential
\begin{equation}
U_{\rm eff}=U(r)+U_1(r)l_z \sigma_z
\label{eq:SO2D}
\end{equation}
in the presence of Rashba SO interaction.
$U_1(r)=-\lambda U'(r)/r$ has the same sign as $U(r)$ if
$|U(r)|$ is a monotonically decreasing function of $r$ and
$\lambda>0$. For electrons with $\sigma_z=1$, $U_{\rm eff}=
U(r) + U_1(r) l_z$ and as a result, the scattering
for components of $l_z>0$ ($l_z<0$) is enhanced (suppressed)
by the SO interaction. For electrons with $\sigma_z=-1$,
the effect is opposite. This is the origin of the extrinsic SHE
in two-dimensional electron system.
We showed that the SHE is significantly enhanced by the resonant
scattering when $U(r)$ is attractive and properly tuned.
We proposed a three-terminal spin-filter including a
single antidot.

In the present study, we examine an enhancement of the
``extrinsic SHE'' by resonant tunneling through a
quantum dot (QD) in multi-terminal geometries.
The QD is a well-known device showing a Coulomb
oscillation when the electrostatic potential is tuned
by a gate voltage.~\cite{Kouwenhoven}
The number of electrons is almost fixed by the Coulomb
blockade between the current peaks of the oscillation.
At the current peaks, the resonant tunneling takes
place through discrete energy levels in the QD at low
temperatures of $k_{\rm B}T \ll \Gamma$ with
level broadening $\Gamma$ due to the tunnel coupling to
the leads.
Recently, the SO interaction in QDs of narrow-gap
semiconductors and related phenomena have been investigated
intensively.~\cite{Igarashi,Fasth,Pfund,Takahashi,Kanai,Deacon,Schroer,Golovach,Nowack,
Nadj-Perge1,Nadj-Perge2,Nadj-Perge3}
We consider a situation in which a QD with SO interaction
is connected to $N$ ($\ge 2$) external leads via
tunnel barriers.
We use the term SHE in the following meaning:
When an unpolarized current is injected
to the QD from a lead (lead S), polarized currents are ejected to
the other leads [D1,$\dots$,D$(N-1)$]. In other words, the QD works
as a spin filter. We assume that the SO interaction is present only
in the QD and that the average of level spacing in the QD is comparable
to the level broadening $\Gamma$ ($\sim 1$ $\mathrm{meV}$),
in accordance with experimental situations.~\cite{Igarashi}
Thus the transport takes place through
single or a few energy levels in the QD around the Fermi level
$\varepsilon_{\rm F}$ in the leads.
The strength of SO interaction $\Delta_\text{SO}$
[absolute value of ${\bm h}_{\rm SO}$ in Eq.\ (\ref{eq:offHso})]
is approximately $0.1 \sim 0.2$ meV
for InAs QDs~\cite{Fasth,Pfund,Takahashi,Kanai}
and $0.23$ meV for InSb QDs.~\cite{Nadj-Perge3}

Our purpose is to elucidate the mechanism of SHE at a
QD with discrete energy levels. Consider an electron with spin-up or
-down injected to the QD from lead D1 (electric current flows from
the QD to lead D1). The SO interaction in the QD mixes
a few energy levels around $\varepsilon_{\rm F}$ in a spin-dependent
way [a rotation in the pseudo-spin space of the levels; see
Eq.\ (\ref{eq:Hdot})], whereas the tunnel coupling to lead
D2 mixes the levels differently in a spin-independent way.
The interference between the mixings results in the spin-polarized
electron going out to lead S.
To simply clarify the spin-dependent transport processes, 
we neglect the electron-electron interaction.
We focus on the current peaks of the Coulomb oscillation where
the interaction is not qualitatively important.

First, we examine a two-level QD as a minimal model and present
an analytical expression for the spin-dependent conductance.
We assume single conduction channel in each of $N$ leads.
In the absence of magnetic field, we show that three or more leads
($N \ge 3$) are required to generate the spin-polarized current.
We observe a large spin polarization by the resonant tunneling
at the current peak when the spacing $\Delta$ between the two levels
in the QD is smaller than $\Gamma$. Although the SHE at a QD seems
quite different from the SHE by an impurity potential, the
condition of $\Delta < \Gamma$ would correspond to the degeneracy
for the virtual bound states with $\pm l_z$ [see Eq.\ (\ref{eq:SO2D})].
The preliminary results of this part in the present paper were
published in our previous paper.~\cite{EYjpsj2}

Second, we analyze the transport through the two-level QD
in a weak magnetic field.
The orbital magnetization is taken into account to the first
order of magnetic field,
whereas the Zeeman effect is neglected.
We find the creation of spin-polarized current in a conventional
geometry of two-terminated QD ($N=2$) with finite magnetization $b$
[see Eq.\ (\ref{eq:offBl}); $b \sim \hbar \omega_{\rm c}$ with cyclotron
frequency $\omega_{\rm c}=|e|B/m^*$] and
enhancement of the polarization when $|b|$ is comparable to the
strength of the SO interaction $\Delta_{\rm SO}$ (magnetic
field of $B \sim 40$ mT).
This is ascribable to the interference between the spin-dependent
mixing of energy levels in the QD by the SO interaction and
spin-independent one by the orbital magnetization.

Finally, our analytical results for the two-level QD are confirmed by
numerical study on the QD with several energy levels. A QD with
tunnel barriers to $N$ leads is modeled on a two-dimensional
tight-binding model.
We observe spin-polarized currents for $N=3$ ($N=2$) in the absence
(presence) of magnetic field. The spin polarization is markedly
enhanced at the current peaks when a few energy levels are close to
each other around $\varepsilon_{\rm F}$.

We make some comments here. (i) 
Previous theoretical
papers~\cite{Kiselev3,Bardarson,Krich1,Krich2}
concerned the spin-current generation in a mesoscopic
region, or an open QD with no tunnel barriers,
in which many energy levels in the QD participate
in the transport. Since we are interested in the resonant
tunneling through one or two discrete levels in the QD,
our situation is different from that in the papers.

(ii)
The present work indicates a QD spin filter in multi-terminal
(two-terminal) geometries without (with) magnetic field
although we emphasize the fundamental aspect of the mechanism
for the SHE at a QD. Note that our spin filter works only at
low temperatures since the SHE stems from the coherent
transport processes through the QD.
Other spin filters were proposed using semiconductor
nanostructures with SO interaction, e.g., 
three- or four-terminal devices related to
the SHE,~\cite{EYjpsj1,YEprb,Bulgakov,Kiselev1,Kiselev2,Pareek,Yamamoto,Nikolic}
a triple-barrier tunnel diode,~\cite{3diode}
quantum point contact,~\cite{EKHjpsj,Silvestrov}
and a three-terminal device for the Stern-Gerlach experiment
using a nonuniform SO interaction.~\cite{3termSG}

(iii)
We do not consider the electron-electron interaction
in the present paper focusing on the current peaks of the
Coulomb oscillation.
In the Coulomb blockade regimes between the current peaks,
the electron-electron interaction plays
a crucial role. We examined the many-body resonance induced
by the Kondo effect in the blockade regime with spin 1/2
in the multi-terminated QD. We showed the generation of
largely polarized current in the presence of the SU(4) Kondo
effect when the level spacing is less than the Kondo
temperature.~\cite{EYjpsj2}
We also mention that an enhancement of SHE by the resonant
scattering or Kondo resonance was examined for metallic systems
with magnetic impurities.~\cite{Fert,Fert2,Guo}

The organization of the present paper is as follows.
In Sec.\ II, we explain a model of two-level QD connected
to $N$ external leads.
Section III presents the analytical expressions for the
spin-dependent conductance using the model of two-level QD.
In Sec.\ IV, we study a generalized situation in which a QD
with many energy levels is connected to $N$ leads through
tunnel barriers. We make a two-dimensional tight-binding model to
describe the situation and perform a numerical study.
The last section (Sec.\ V) is devoted to the conclusions
and discussion.

\begin{figure*}
\begin{center}
\includegraphics[width=15cm]{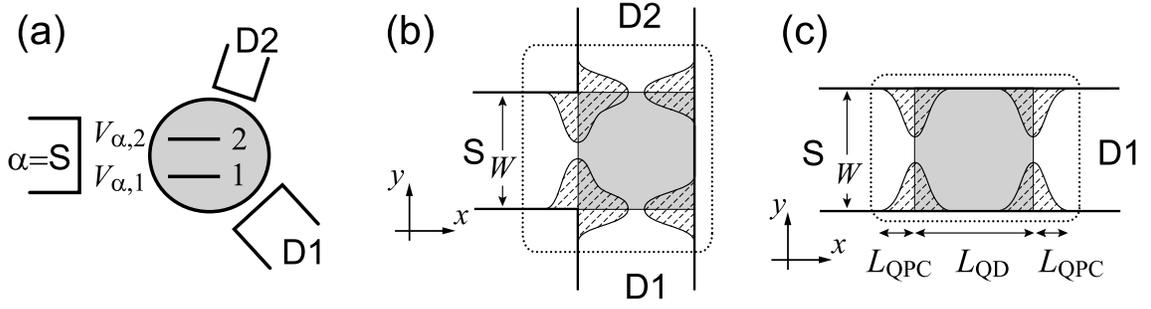}
\end{center}
\caption{
Models for a quantum dot (QD) connected to $N$ ($\geq 2$) leads.
When an unpolarized current is injected to the QD from lead S,
polarized currents are ejected to leads D1 to D$(N-1)$.
(a) A QD with two energy levels. The tunnel coupling
between level $j$ ($=1,2$) in the QD and lead $\alpha$ is
denoted by $V_{\alpha,j}$.
(b) A QD (shaded square region of $W \times W$ in area) connected
to three leads (quantum wires of $W$ in width) via quantum point
contacts. (c) A QD connected to two leads via quantum point
contacts in the presence of magnetic field.
Models in (b) and (c) are represented on a tight-binding model
by discretizing the two-dimensional space ($xy$ plane).
}
\label{fig:model}
\end{figure*}

\section{MODEL OF TWO-LEVEL QUANTUM DOT}
\label{sec:model}

In this section, we explain our model depicted in Fig.\ 1(a),
in which a two-level QD is connected to $N$ external leads.

We start from a QD with SO interaction and magnetic field $\bm{B}$
in general. The electronic state in the QD is described by
the Hamiltonian
\begin{eqnarray}
H_{\rm dot}^{(0)} &=& \frac{(\bm{p}-e\bm{A})^2}{2m^*}
+ U(\bm{r}) + H_{\rm SO}(\bm{B})
\label{eq:generalH} \\
&\simeq & \frac{\bm{p}^2}{2m^*} + U (\bm{r})
+ \frac{|e| \hbar}{2m^*} \bm{B} \cdot \bm{l} + H_{\rm SO},
\label{eq:weakH}
\end{eqnarray}
where $U(\bm{r})$ is the confining potential of the QD,
$m^*$ is the effective mass of conduction electrons
($m^*/m_0=0.024$ in InAs with $m_0$ being the electron mass
in the vacuum),
and $\bm{A}=(\bm{B} \times \bm{r})/2$ is the vector
potential. Assuming a weak magnetic field, we neglect
the term of $\bm{A}^2$ and Zeeman effect.
For the SO interaction, $H_{\rm SO}$ can be the Rashba
and/or Dresselhaus interactions in Eqs.\ (\ref{eq:SORashba})
and (\ref{eq:Dresselhaus}).
Although $\bm{p}$ in $H_{\rm SO}$ should be replaced by
$(\bm{p}-e\bm{A})$ in the presence of magnetic field,
the terms of $\bm{A}$ in $H_{\rm SO}(\bm{B})$ can be disregarded
in the case of weak magnetic field
(see Appendix A).

The eigenenergies of $\bm{p}^2/(2m^*) + U (\bm{r})$ form
a set of discrete energy levels $\{ \varepsilon_i \}$.
We examine the situation in which two energy levels,
$\varepsilon_1$ and $\varepsilon_2$,
are relevant to the transport. The other levels are located
so far from the two levels that the mixing by $H_{\rm SO}$ or
$|e| \hbar \bm{B} \cdot \bm{l}/(2m^*)$ can be neglected.

The wavefunctions of the states, $\langle {\bm r} |1 \rangle$ and
$\langle {\bm r} |2 \rangle$, can be real since they are
eigenstates of real operator, $\bm{p}^2/(2m^*) + U (\bm{r})$.
Since the orbital part in $H_{\rm SO}$
is a pure imaginary operator, it has off-diagonal elements only;
\begin{equation}
\langle 2| H_{\rm SO} | 1 \rangle =
{\rm i} {\bm h}_{\rm SO} \cdot {\bm \sigma}/2
\label{eq:offHso}
\end{equation}
with ${\bm h}_{\rm SO}={\bm h}_{\rm RSO}+{\bm h}_{\rm DSO}$.
${\rm i} {\bm h}_{\rm RSO}= (2\lambda/\hbar) 
\langle 2 | ({\bm p} \times {\bm \nabla} U) | 1 \rangle$
in the case of Rashba interaction, whereas
${\rm i} h_{{\rm DSO},x}=(2\lambda^\prime /\hbar)
\langle 2 | p_x (p_y^2 - p_z^2) | 1 \rangle$, etc.,
in the case of Dresselhaus interaction.
For the same reason,
$\langle 1| \bm{B} \cdot \bm{l} | 1 \rangle =
 \langle 2| \bm{B} \cdot \bm{l} | 2 \rangle =0$
and
\begin{equation}
\frac{|e| \hbar}{m^*} \langle 2| \bm{B} \cdot \bm{l} | 1 \rangle =
{\rm i} b/2.
\label{eq:offBl}
\end{equation}
We estimate the value of $|b|$ to be
$|e| \hbar B/m^*=\hbar \omega_{\rm c}$,
where $\omega_{\rm c}=|e|B/m^*$ is the cyclotron frequency.
When $\hbar \omega_{\rm c}=0.2$ meV ($\simeq \Delta_{\rm SO}$),
the corresponding magnetic field is $B=40$ mT in the case of InAs.
If the quantization axis of spin is taken in the
direction of ${\bm h}_{\rm SO}$,
the Hamiltonian in the QD reads
\begin{equation}
H_{\rm dot}= \sum_{\sigma = \pm}
(d_{1,\sigma}^\dagger ,\ d_{2,\sigma}^\dagger )
\left(
\bar{\varepsilon} - \frac{\Delta}{2}
\tau_z  +  \frac{b + \sigma \Delta_\text{SO}}{2} \tau_y
\right) \left( \begin{array}{c}
d_{1,\sigma} \\
d_{2,\sigma}
\end{array} \right),
\label{eq:Hdot}
\end{equation}
where $d_{j,\sigma}^{\dagger}$ and $d_{j,\sigma}$ are the
creation and annihilation operators of an electron with
orbital $j$ and spin $\sigma$, respectively.
$\bar{\varepsilon} =(\varepsilon_1 + \varepsilon_2 )/2$,
$\Delta =\varepsilon_2 -\varepsilon_1$,
and $\Delta_\text{SO}=|{\bm h}_\text{SO}|$.
The Pauli matrices, $\tau_y$ and $\tau_z$, 
are introduced for the pseudo-spin representing level
$1$ or $2$ in the QD.
Note that the
Hamiltonian in Eq.\ (\ref{eq:Hdot}) yields the energy levels
of $\bar{\varepsilon} \pm
\sqrt{\Delta^2+(b+\sigma \Delta_\text{SO})^2}/2$ for
$\sigma=+$ or $-$ in an isolated QD;
the Kramers degeneracy holds only with $b=0$.
Although the average of level spacing in a QD
is assumed to be $\delta \sim 1$ meV, the spacing $\Delta$
between a specific pair of levels fluctuates around $\delta$.
$\Delta$ is fixed while the electrostatic potential, and
hence the mean energy level $\bar{\varepsilon}$,
is changed by tuning the gate voltage.

The state $|j \rangle$ in the QD is connected to lead
$\alpha$ by tunnel coupling, $V_{\alpha, j}$ ($j=1,2$),
which is real. The tunnel Hamiltonian is
\begin{eqnarray}
H_{\rm T} & = &
\sum_{j=1}^2 \sum_{\alpha=1}^N \sum_{k,\sigma}
\left( V_{\alpha, j} d_{j,\sigma}^{\dagger}a_{\alpha k,\sigma}+
\text{h.c}.\right)
\nonumber \\
& = &
\sum_{\alpha=1}^N \sum_{k,\sigma}
V_{\alpha} \left[ \left( e_{\alpha,1} d_{1,\sigma}^{\dagger}+
e_{\alpha,2} d_{2,\sigma}^{\dagger} \right) a_{\alpha k,\sigma}+
\text{h.c}.\right],
\label{eq:HT}
\end{eqnarray}
where $a_{\alpha k,\sigma}$ annihilates an electron with
state $k$ and spin $\sigma$ in lead $\alpha$.
$V_{\alpha}=\sqrt{(V_{\alpha, 1})^2+(V_{\alpha, 2})^2}$
and $e_{\alpha,j}=V_{\alpha, j}/V_{\alpha}$. We
introduce a unit vector,
${\bm e}_{\alpha}=(e_{\alpha,1},e_{\alpha,2})^{\rm T}$.
$V_{\alpha}$ is controllable by electrically tuning
the tunnel barrier, whereas ${\bm e}_{\alpha}$ is
determined by the wavefunctions
$\langle {\bm r} | 1 \rangle$ and
$\langle {\bm r} | 2 \rangle$ in the QD and hardly
controllable for a given current peak.
It should be mentioned that
$\{ {\bm e}_{\alpha} \}$ and $\Delta$ vary
from peak to peak in the Coulomb oscillation.
We can choose a peak with appropriate parameters
for the SHE in experiments.

We assume a single channel of conduction electrons
in the leads. The total Hamiltonian is
\begin{equation}
H=\sum_{\alpha=1}^N \sum_{k,\sigma} \varepsilon_k
  c_{\alpha k,\sigma}^{\dagger} c_{\alpha k,\sigma}
  +H_\text{dot}+H_\text{T}.
\label{eq:Hamiltonian}
\end{equation}

The strengths of tunnel couplings to lead $\alpha$
are characterized by the level broadening,
$\Gamma_\alpha =\pi \nu_\alpha (V_{\alpha})^2$,
where $\nu_\alpha$ is the density of states in the lead.
We also introduce a matrix of
$\hat{\Gamma}=\sum_{\alpha} \hat{\Gamma}_{\alpha}$
with
\begin{equation}
\hat{\Gamma}_{\alpha}
=
\Gamma_\alpha
\left( \begin{array}{cc}
(e_{\alpha, 1})^2 & e_{\alpha, 1}e_{\alpha, 2}
\\
e_{\alpha, 1}e_{\alpha, 2} & (e_{\alpha, 2})^2
\end{array}
\right).
\label{eq:Gammamatrx}
\end{equation}

\begin{figure*}
\begin{center}
\includegraphics[width=12cm]{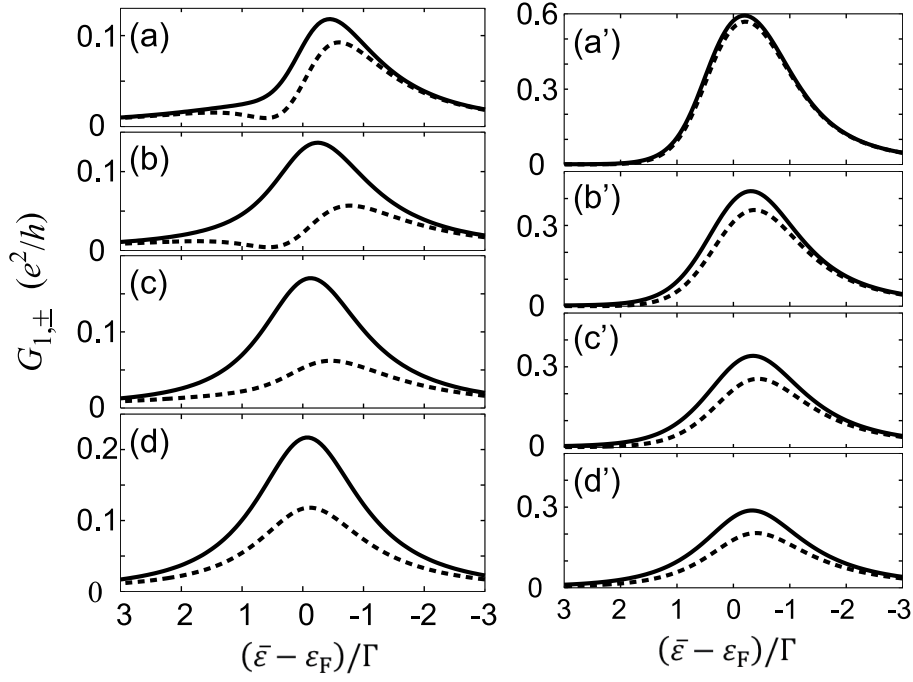}
\end{center}
\caption{
Spin-dependent conductance $G_{1, \pm}$ in the model of
two-level quantum dot in the three-terminal geometry,
as a function of mean energy level,
$\bar{\varepsilon} = (\varepsilon_1 + \varepsilon_2 )/2$.
No magnetic field is applied.
Solid (broken) lines indicate the conductance
$G_{1,+}$ ($G_{1,-}$) for spin $\sigma=+1$ ($-1$)
in the direction of ${\bm h}_{\rm SO}$ (see Sec.\ II).
The level spacing in the quantum dot is
$\Delta=\varepsilon_2 -\varepsilon_1=0.2\Gamma$
(left panels) and $\Gamma$ (right panels).
The level broadening by the tunnel coupling to
leads S and D$1$ is
$\Gamma_{\rm S} = \Gamma_{{\rm D}1} \equiv \Gamma$
($e_{{\rm S}, 1}/e_{{\rm S}, 2} = 1$, 
$e_{{\rm D}1, 1}/e_{{\rm D}1, 2} = -2/3$),
whereas that to lead D$2$ is
(a) $\Gamma_{{\rm D}2} =0.1\Gamma$,
(b) $0.5\Gamma$, (c) $\Gamma$, and (d) $2\Gamma$
($e_{{\rm D}2, 1}/e_{{\rm D}2, 2} = 2$).
The strength of spin-orbit interaction is fixed at
$\Delta_{\rm SO} =0.2\Gamma$.
}
\label{fig:LM3cond}
\end{figure*}

An unpolarized current is injected into the QD from a
source lead ($\alpha=$S) and output to other leads
[D$n$; $n=1,\cdots,(N-1)$]. The electrochemical potential
for electrons in lead S is lower than that in the other leads
by $|e|V_\text{bias}$.
The transport through a QD in the multi-terminal geometry
can be formulated following the paper by Meir and
Wingreen,~\cite{MeirWingreen} just as in the two-terminal geometry.
The current with spin $\sigma=\pm$ from lead $\alpha$ to
the QD is written as
\begin{equation}
I_{\alpha,\sigma}=\frac{\text{i}e}{\pi \hbar} \int d\varepsilon
\text{Tr} \left\{
\hat{\Gamma}_{\alpha}
\left[f_{\alpha}(\varepsilon)
\left(\hat{G}^\text{r}_{\sigma}-\hat{G}^\text{a}_{\sigma} \right)+
\hat{G}^{<}_{\sigma} \right]
\right\},
\label{eq:current}
\end{equation}
where $\hat{G}^\text{r}_{\sigma}$,
$\hat{G}^\text{a}_{\sigma}$,
and $\hat{G}^{<}_{\sigma}$ are the retarded, advanced,
and lesser Green functions in the QD,
respectively, in $2\times 2$ matrix form
in the pseudo-spin space.
$f_{\alpha}(\varepsilon)$
is the Fermi distribution function in lead $\alpha$.

Although the current formula in Eq.\ (\ref{eq:current}) is
applicable in the presence of electron-electron
interaction in the QD, it is simplified in its absence.
Then, $\hat{G}^\text{r}_{\sigma}-\hat{G}^\text{a}_{\sigma}
=-2\text{i} \hat{G}^\text{r}_{\sigma} \hat{\Gamma}
\hat{G}^\text{a}_{\sigma}$ and
$\hat{G}^{<}_{\sigma}=2\text{i} \hat{G}^\text{r}_{\sigma}
(\sum_{\alpha} \hat{\Gamma}_{\alpha} f_{\alpha})
\hat{G}^\text{a}_{\sigma}$. The substitution of
these relations into Eq.\ (\ref{eq:current}) yields
\[
I_{\text{D}n,\sigma}=\frac{4e}{h} \int d\varepsilon
\left[ f_\text{D}(\varepsilon)-f_\text{S}(\varepsilon) \right]
\text{Tr} \left( \hat{G}^\text{a}_{\sigma}
\hat{\Gamma}_{\text{D}n} \hat{G}^\text{r}_{\sigma}
\hat{\Gamma}_\text{S} \right),
\]
where
$f_{\text{D}n}(\varepsilon) \equiv f_\text{D}(\varepsilon)$.
At $T=0$, the conductance into lead D$n$ with spin $\sigma$
is given by
\begin{equation}
G_{n,\sigma}=
\left.
-\frac{\text{d}I_{\text{D}n,\sigma}}{\text{d}V_\text{bias}}
\right|_{V_\text{bias} = 0}
=\frac{4e^2}{h} \text{Tr}
\left[ \hat{G}^\text{a}_{\sigma}(\varepsilon_\text{F})
\hat{\Gamma}_{\text{D}n}
\hat{G}^\text{r}_{\sigma}(\varepsilon_\text{F})
\hat{\Gamma}_{\text{S}} \right],
\label{eq:conductance0}
\end{equation}
where the QD Green function is
\begin{equation}
\hat{G}^\text{r}_{\pm}(\varepsilon) =
\left[
\left( \begin{array}{cc}
\varepsilon-\varepsilon_\text{d} + \frac{\Delta}{2}
 & \text{i} \frac{b \pm \Delta_\text{SO}}{2}
\\
- \text{i} \frac{b \pm \Delta_\text{SO}}{2} &
\varepsilon-\varepsilon_\text{d} - \frac{\Delta}{2}
\end{array}
\right)
+\text{i} \hat{\Gamma}
\right]^{-1}.
\label{eq:G0}
\end{equation}

\section{ANALYTICAL RESULTS}
\label{sec:Results}

We analyze the model of two-level QD, introduced in the
previous section. We show analytical expressions for the spin-dependent
conductance in the absence and presence of magnetic field, respectively.

\subsection{In absence of magnetic field}

We begin with the case of $b=0$, or in the absence of magnetic field.
From Eqs.\ (\ref{eq:conductance0}) and
(\ref{eq:G0}), we obtain
\begin{eqnarray}
G_{n,\sigma} & = & \frac{e^2}{h}
\frac{4\Gamma_\text{S} \Gamma_{\text{D}n}}{|D|^2}
\left[ g_{n}^{(1)} + g_{n,\sigma}^{(2)} \right],
\label{eq:conductance}
\\
g_{n}^{(1)} & = &
\Biggl[
\left(
\varepsilon_\text{F}-\bar{\varepsilon}-\frac{\Delta}{2}
\right)
e_{\text{D}n,1} e_{\text{S},1}
\nonumber \\
& &
+
\left(
\varepsilon_\text{F}-\bar{\varepsilon}+\frac{\Delta}{2}
\right)
e_{\text{D}n,2} e_{\text{S},2}
\Biggr]^2,
\label{eq:g1}
\\
g_{n,\pm}^{(2)} & = &
\Biggl[
\pm \frac{\Delta_\text{SO}}{2}
({\bm e}_\text{S} \times {\bm e}_{\text{D}n})_z
\nonumber \\
& &
+
\sum_{\alpha} \Gamma_{\alpha}
({\bm e}_{\text{D}n} \times {\bm e}_{\alpha})_z
({\bm e}_\text{S} \times {\bm e}_{\alpha})_z
\Biggr]^2,
\label{eq:g2noB}
\end{eqnarray}
where $D$ is the determinant of
$[\hat{G}^\text{r}_{\sigma}(\varepsilon_\text{F})]^{-1}$
in Eq.\ (\ref{eq:G0}), which is independent of $\sigma$.
$({\bm a} \times {\bm b})_z = a_1 b_2 - a_2 b_1$.

Let us consider two simple cases. (I)
When
$\Delta \gg \Gamma_{\alpha}$ and $\Delta_\text{SO}$, 
$G_{n,\sigma}$ consists of
two Lorentzian peaks as a function of
$\bar{\varepsilon}$, reflecting the resonant
tunneling through one of the energy levels,
$\varepsilon_{1,2}=
\bar{\varepsilon} \mp \Delta/2$:
\begin{equation}
G_{n,\sigma} \approx \frac{4 e^2}{h}
\Gamma_\text{S} \Gamma_{\text{D}n}
\sum_{j=1,2}
\frac{(e_{\text{D}n,j} e_{\text{S},j})^2}
{(\varepsilon_j-\varepsilon_\text{F})^2+(\Gamma_{jj})^2}.
\label{eq:G-largeD}
\end{equation}
Here, $\Gamma_{jj}=\sum_{\alpha} \pi \nu_{\alpha} (V_{\alpha,j})^2$
is the broadening of level $j$
($jj$ component of matrix $\hat{\Gamma}$).
In this case,
the spin-polarized current [$\propto (G_{n,+}-G_{n,-})$] is
very small. $\Delta$ should be comparable to
or smaller than the level broadening
to observe a considerable spin current.
(II) In a two-terminated QD ($N=2$), the second term in
$g_{n,\pm}^{(2)}$ vanishes.
Since $g_{n,+}^{(2)}=g_{n,-}^{(2)}$,
no spin-polarized current is generated.~\cite{com1}
Three or more leads are required to generate a spin-polarized
current, as pointed out by other groups.~\cite{Krich1,Zhai,Kiselev3}

\begin{figure}
\begin{center}
\includegraphics[width=8.5cm]{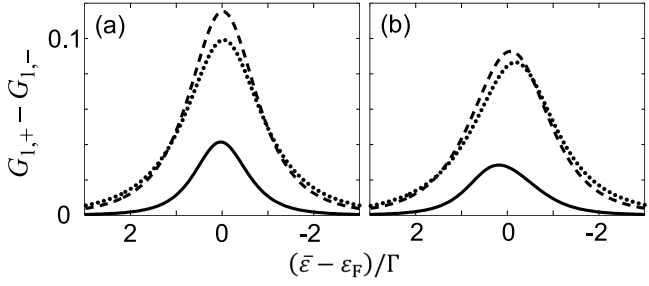}
\end{center}
\caption{
Spin-polarized conductance, $G_{1,+}-G_{1,-}$,
in the model of two-level quantum dot in the three-terminal
geometry, as a function of $\bar{\varepsilon}$.
The level spacing in the quantum dot is
(a) $\Delta=\varepsilon_2 -\varepsilon_1=0.2\Gamma$
and (b) $\Gamma$.
$\Gamma_{{\rm D}2} =0.1\Gamma$ (solid line),
$\Gamma$ (broken line), and $2\Gamma$ (dotted line).
The other parameters are the same as in Fig.\ 2.
}
\label{fig:LM3polar}
\end{figure}

We examine $G_{1,\pm}$ in the three-terminated
system ($N=3$) in the rest of this subsection. Then
$g_{1,\pm}^{(2)} =
[\pm (\Delta_\text{SO}/2)
({\bm e}_\text{S} \times {\bm e}_{\text{D}1})_z +
\Gamma_{\text{D}2}
({\bm e}_{\text{D}1} \times {\bm e}_{\text{D}2})_z
({\bm e}_\text{S} \times {\bm e}_{\text{D}2})_z
]^2$.
We exclude specific situations in which two out of
${\bm e}_\text{S}$,
${\bm e}_{\text{D}1}$, and ${\bm e}_{\text{D}2}$ are
parallel to each other.
The conditions for a largely spin-polarized current are
as follows:
(i) $\Delta \lesssim$ (level broadening),
as mentioned above. Two levels in the QD should
participate in the transport.
(ii) The Fermi level in the leads is close to the energy
levels in the QD,
$\varepsilon_\text{F} \approx \bar{\varepsilon}$
(resonant condition).
(iii) The level broadening by the tunnel coupling to
lead D$2$, $\Gamma_{\text{D}2}$, is comparable to
the strength of SO interaction $\Delta_\text{SO}$.

Figures 2 and 4 show two typical results of
the conductance $G_{1,\pm}$ as a function of
$\bar{\varepsilon}$.
In $g_{1}^{(1)}$, $e_{\text{D}1,1} e_{\text{S},1}$ and
$e_{\text{D}1,2} e_{\text{S},2}$ have different
(same) signs in Fig.\ 2 (Fig.\ 4). Therefore,
$g_{1}^{(1)}=0$ has no solution (a solution) in
$-\Delta/2 <
\bar{\varepsilon}-\varepsilon_\text{F}
< \Delta/2$.

\begin{figure}
\begin{center}
\includegraphics[width=8.5cm]{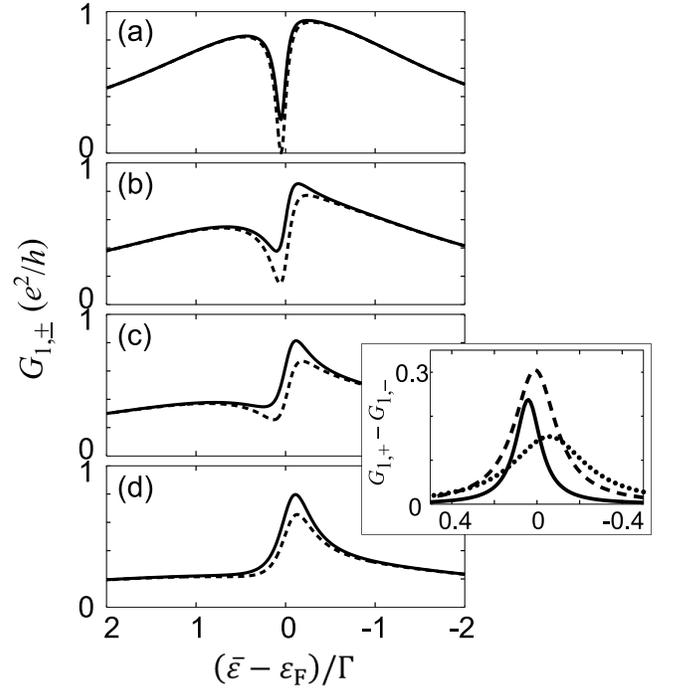}
\end{center}
\caption{
Spin-dependent conductance $G_{1, \pm}$ in the model of
two-level quantum dot in the three-terminal geometry,
as a function of mean energy level,
$\bar{\varepsilon} = (\varepsilon_1 + \varepsilon_2 )/2$.
No magnetic field is applied.
Solid (broken) lines indicate the conductance
$G_{1,+}$ ($G_{1,-}$) for spin $\sigma=+1$ ($-1$)
in the direction of ${\bm h}_{\rm SO}$ (see Sec.\ II).
The level spacing in the quantum dot is
$\Delta=\varepsilon_2 -\varepsilon_1=0.5\Gamma$.
The level broadening by the tunnel coupling to
leads S and D$1$ is
$\Gamma_{\rm S} = \Gamma_{{\rm D}1} \equiv \Gamma$
($e_{{\rm S}, 1}/e_{{\rm S}, 2} = 1$, 
$e_{{\rm D}1, 1}/e_{{\rm D}1, 2} = 2/3$),
whereas that to lead D$2$ is
(a) $\Gamma_{{\rm D}2} =0.1\Gamma$,
(b) $0.5\Gamma$, (c) $\Gamma$, and (d) $2\Gamma$
($e_{{\rm D}2, 1}/e_{{\rm D}2, 2} = 2$).
The strength of spin-orbit interaction is fixed at
$\Delta_{\rm SO} =0.2\Gamma$.
Inset: Spin-polarized conductance, $\propto G_+ - G_-$,
as a function of $\bar{\varepsilon}$.
$\Gamma_{{\rm D}2} =0.1\Gamma$ (solid line),
$0.5\Gamma$ (broken line), and $2\Gamma$ (dotted line).
}
\label{fig:LM3phaselapse}
\end{figure}

In Fig.\ 2, the conductance shows a single peak.
We set $\Gamma_\text{S}=\Gamma_{\text{D}1} \equiv \Gamma$
and change $\Gamma_{\text{D}2}$ from (a) $0.1\Gamma$ to
(d) $2\Gamma$.
When $\Delta=0.2 \Gamma$ (left panels),
we observe a large spin polarization around the current
peak, which clearly indicates an enhancement of the SHE
by the resonant tunneling [conditions (i) and (ii)].
With increasing $\Gamma_{\text{D}2}$, the spin
current increases first, takes a maximum in panel (c),
and then decreases [condition (iii)].
This means that the SHE is tunable by changing the tunnel
coupling. When $\Delta=\Gamma$ (right panels), the SHE is
less effective; spin polarization of
$P=(G_{1,+}-G_{1,-})/(G_{1,+}+G_{1,-})$ around the current
peak is smaller than in the case of $\Delta=0.2 \Gamma$.
However, a value of spin-polarized conductance, $G_{n,+}-G_{n,-}$, is
still large, as depicted in Fig.\ 3.

\begin{figure*}
\begin{center}
\includegraphics[width=12cm]{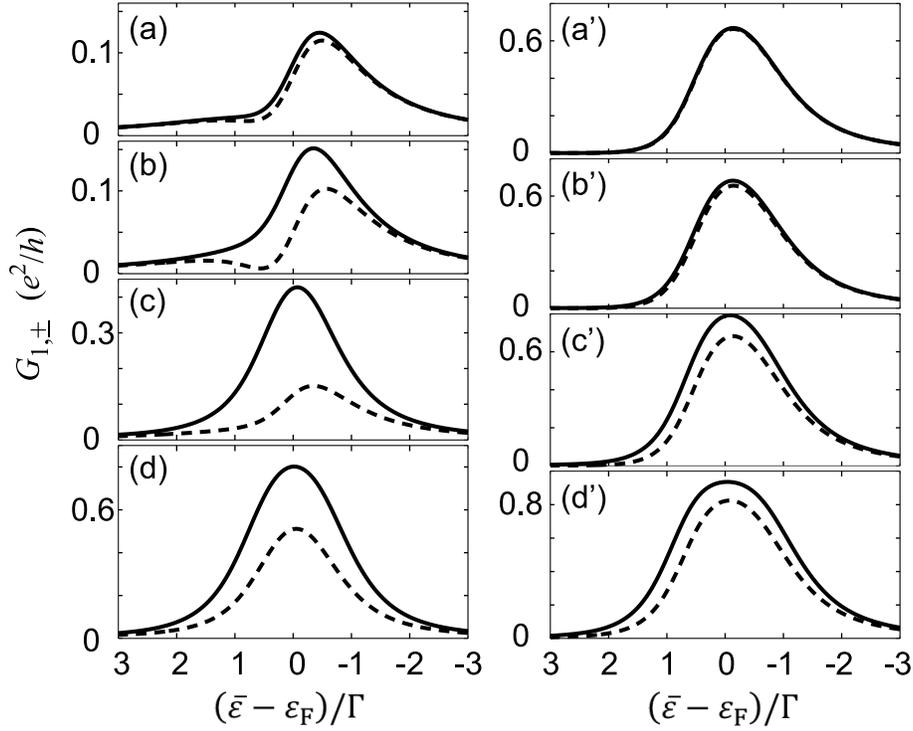}
\end{center}
\caption{
Spin-dependent conductance $G_{1, \pm}$ in the model of
two-level quantum dot in the two-terminal geometry,
as a function of mean energy level,
$\bar{\varepsilon} = (\varepsilon_1 + \varepsilon_2 )/2$,
in the presence of magnetic field.
Solid (broken) lines indicate the conductance
$G_{1,+}$ ($G_{1,-}$) for spin $\sigma=+1$ ($-1$)
in the direction of ${\bm h}_{\rm SO}$ (see Sec.\ II).
The level spacing in the quantum dot is
$\Delta=\varepsilon_2 -\varepsilon_1=0.2\Gamma$
(left panels) and $\Gamma$ (right panels).
The level broadening by the tunnel coupling to
leads S and D$1$ is
$\Gamma_{\rm S} = \Gamma_{{\rm D}1} \equiv \Gamma$
($e_{{\rm S}, 1}/e_{{\rm S}, 2} = 1$, 
$e_{{\rm D}1, 1}/e_{{\rm D}1, 2} = -2/3$).
The orbital magnetization is
(a) $b =0.02\Gamma$, (b) $0.1\Gamma$, (c) $0.5\Gamma$,
and (d) $\Gamma$.
The strength of spin-orbit interaction is fixed at
$\Delta_{\rm SO} =0.2\Gamma$.
}
\label{fig:LMBcond}
\end{figure*}

In Fig.\ 4, the conductance $G_{1,\pm}$ shows a dip
at $\bar{\varepsilon} \approx \varepsilon_\text{F}$
for small $\Gamma_{\text{D}2}$.
The conductance dip is caused by the destructive
interference between propagating waves through
two orbitals in the QD.
In the two-terminated QD without SO interaction
($\Gamma_{\text{D}2}=\Delta_{\rm SO}=0$),
the conductance $G_{1,\sigma} \propto g_{1}^{(1)}$
would completely vanish at the dip, where the
``phase lapse'' of the transmission phase takes
place.~\cite{Karrasch}
As seen in Fig.\ 4, the conductance dip changes to
a peak with increasing $\Gamma_{\text{D}2}$ in the
three-terminated QD. The SO interaction makes a
large difference between $G_{1,+}$ and $G_{1,-}$
around the dip or peak, similarly to the case in Fig.\ 2.
The spin-polarized conductance, $G_{1,+}-G_{1,-}$, shows
a large peak there, as seen in the inset in Fig.\ 4.
In Fig.\ 4(a) with $\Gamma_{\text{D}2}=0.1\Gamma$,
we find that the spin polarization of
$P=(G_{1,+}-G_{1,-})/(G_{1,+}+G_{1,-})$ is close to unity
around the dip since $G_{1,-}$ is almost zero.

\subsection{In presence of magnetic field}

Now we discuss the case with magnetic field: $b \ne 0$.
The conductance into lead D$n$ with spin $\sigma$ is
\begin{equation}
G_{n,\sigma} = \frac{e^2}{h}
\frac{4\Gamma_\text{S} \Gamma_{\text{D}n}}{|D_{\sigma}|^2}
\left[ g_{n}^{(1)} + g_{n,\sigma}^{(2)} \right],
\label{eq:conductance-B}
\end{equation}
where $g_{n}^{(1)}$ is the same as that in Eq.\ (\ref{eq:g1}),
whereas
\begin{eqnarray}
g_{n,\pm}^{(2)} & = &
\Biggl[
\frac{b \pm \Delta_\text{SO}}{2}
({\bm e}_\text{S} \times {\bm e}_{\text{D}n})_z
\nonumber \\
& &
+
\sum_{\alpha} \Gamma_{\alpha}
({\bm e}_{\text{D}n} \times {\bm e}_{\alpha})_z
({\bm e}_\text{S} \times {\bm e}_{\alpha})_z
\Biggr]^2.
\label{eq:g2-B}
\end{eqnarray}
The determinant of
$[\hat{G}^\text{r}_{\sigma}(\varepsilon_\text{F})]^{-1}$,
$D_{\sigma}$, depends on $\sigma$ in this case.

In contrast to the case of $b=0$, we observe the
spin-dependent transport in a conventional geometry
of two-terminated QD ($N=2$). Then
$g_{1,\pm}^{(2)} = (b \pm \Delta_{\rm SO})^2
({\bm e}_{\rm S} \times {\bm e}_{{\rm D}1})_z^2 /4$.
We expect a large spin polarization when
(iii') $b$ and $\Delta_{\rm SO}$ are comparable to each other,
besides conditions (i) and (ii) in the previous sebsection are
satisfied.

We focus on the two-terminated QD ($N=2$) in this subsection.
Figures 5 and 7 exhibit the spin-dependent
conductance $G_{1,\pm}$ as a function of $\bar{\varepsilon}$.
$e_{\text{D}1,1} e_{\text{S},1}$ and
$e_{\text{D}1,2} e_{\text{S},2}$ have different
(same) signs in Fig.\ 5 (Fig.\ 7).
We set $\Gamma_\text{S}=\Gamma_{\text{D}1} \equiv \Gamma$,
whereas the orbital magnetization is gradually increased from
(a) to (d), or from (a) to (c).

\begin{figure}
\begin{center}
\includegraphics[width=8.5cm]{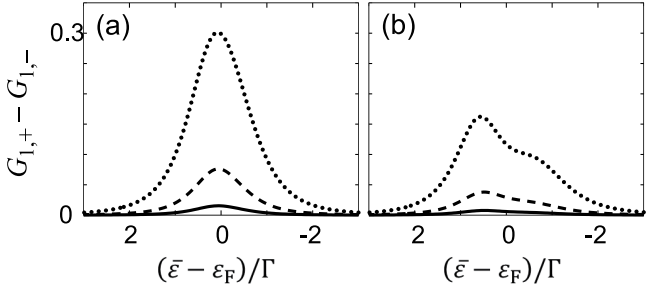}
\end{center}
\caption{
Spin-polarized conductance, $G_{1,+}-G_{1,-}$,
in the model of two-level quantum dot in the two-terminal
geometry, as a function of $\bar{\varepsilon}$,
in the presence of magnetic field.
The level spacing in the quantum dot is
(a) $\Delta=\varepsilon_2 -\varepsilon_1=0.2\Gamma$
and (b) $\Gamma$.
The orbital magnetization $b$ is
$b =0.02\Gamma$ (solid line), $0.1\Gamma$ (broken line),
and $0.5\Gamma$ (dotted line).
The other parameters are the same as in Fig.\ 5.
}
\label{fig:LMBDpolar}
\end{figure}

In Fig.\ 5, the level spacing in the QD is $\Delta=0.2\Gamma$
in the left panels and $\Delta=\Gamma$ in the right panels.
In the absence of magnetic field ($b=0$), we do not observe
the spin-polarized current in the two-terminal geometry,
as discussed in the previous subsection. With an increase in
$b$, the difference between $G_{1,+}$ and $G_{1,-}$ increases,
becomes maximal at $b \sim \Delta_{\rm SO}$,
and decreases [condition (iii')].
The SHE is more prominent for $\Delta=0.2\Gamma$ than for
$\Delta=\Gamma$; the polarization $P$ is larger in the former.

Figure 6 shows the spin-polarized conductance,
$G_{1,+}-G_{1,-}$, as a function of $\bar{\varepsilon}$.
We observe a large value even in the case of $\Delta=\Gamma$
if both the magnetic field and $\bar{\varepsilon}$ are
properly tuned.

In Fig.\ 7, we observe a dip of conductance at
$\bar{\varepsilon} \approx \varepsilon_\text{F}$.
Around the dip, the spin-polarized current is largely
enhanced as shown in the inset.
In Fig.\ 7(a) with $b=0.1\Gamma$, the spin polarization of
$P=(G_{1,+}-G_{1,-})/(G_{1,+}+G_{1,-})$ is close to unity
because $G_{1,-}$ almost vanishes.

\section{NUMERICAL STUDY}

In the previous section, we have presented the analytical
expressions for the spin-dependent conductance for the
model of two-level QD. We have illustrated the generation of
spin-polarized current in three- and two-terminated geometries
in the absence and presence of magnetic field, respectively.
In this section, we perform numerical studies for the
QD with many energy levels to confirm our analytical results.
A QD with tunnel barriers to $N$ leads ($N=2,3$) in Figs.\
1(b) and (c) is modeled on the tight-binding model in the $xy$ plane.

\begin{figure}
\begin{center}
\includegraphics[width=8.5cm]{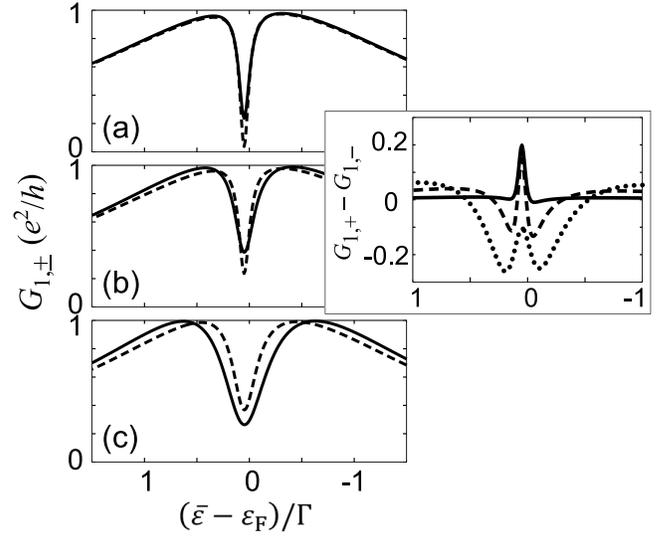}
\end{center}
\caption{
Spin-dependent conductance $G_{1, \pm}$ in the model of
two-level quantum dot in the two-terminal geometry,
as a function of mean energy level,
$\bar{\varepsilon} = (\varepsilon_1 + \varepsilon_2 )/2$,
in the presence of magnetic field.
Solid (broken) lines indicate the conductance
$G_{1,+}$ ($G_{1,-}$) for spin $\sigma=+1$ ($-1$)
in the direction of ${\bm h}_{\rm SO}$ (see Sec.\ II).
The level spacing in the quantum dot is
$\Delta=\varepsilon_2 -\varepsilon_1=0.5\Gamma$.
The level broadening by the tunnel coupling to
leads S and D$1$ is
$\Gamma_{\rm S} = \Gamma_{{\rm D}1} \equiv \Gamma$
($e_{{\rm S}, 1}/e_{{\rm S}, 2} = 1$, 
$e_{{\rm D}1, 1}/e_{{\rm D}1, 2} = 2/3$).
The orbital magnetization is
(a) $b =0.1\Gamma$, (b) $0.5\Gamma$, and (c) $\Gamma$.
The strength of spin-orbit interaction is fixed at
$\Delta_{\rm SO} =0.2\Gamma$.
Inset: Spin-polarized conductance, $\propto G_+ - G_-$,
as a function of $\bar{\varepsilon}$.
$b =0.1\Gamma$ (solid line), $0.5\Gamma$ (broken line),
and $\Gamma$ (dotted line).
}
\label{fig:LMBphaselapse}
\end{figure}

\subsection{Model}
\label{sec:TBmodel}

In Figs.\ 1(b) and (c), $N$ leads connect to a QD
via tunnel barriers.
The $N$ leads are represented by quantum wires of width $W$
with hard-wall potential at the edges.
The electrostatic potential in the QD (shaded square region
of $W \times W$) is changed by $eV_\text{g}$.

The tunnel barriers are described by quantum point contacts (QPCs).
Along a quantum wire in the $x$ direction, the QPC is described by
the potential~\cite{Ando}
\begin{eqnarray}
U(x,y;U_0) &=&
\Bigg\{ \frac{U_0}{2} \left[ 1+\cos \left(
\frac{\pi x}{L_{\rm QPC}} \right) \right] \nonumber \\
& + & \varepsilon_{\rm F} \sum_{\pm} \left(
\frac{y-y_\pm (x)}{W_{\rm QPC}} \right)^2
\theta (y^2 -y_\pm (x)^2) \Bigg\}
\label{eq:QPC}
\end{eqnarray}
at $-L_{\rm QPC} < x < L_{\rm QPC}$, where
\begin{equation}
y_\pm (x)=\pm \frac{W}{4}
\left[ 1-\cos \left( \frac{\pi x}{L_{\rm QPC}} \right)\right]
\end{equation}
and $\theta (t)$ is a step function [$\theta =1$ for
$t>0$, $\theta =0$ for $t<0$].
$U_0$ is the potential height of the saddle point of QPC,
whereas $L_{\rm QPC}$ and $W_{\rm QPC}$ characterize
the thickness and width of the QPC, respectively.
In the QD (shaded square region), the QPC potential is modified
to $U(x,y;U_0 -eV_\text{g}) +eV_\text{g}$.
In Fig.\ 1(b), we cut off the QPC potential at the diagonal lines
of the square to avoid the overlap of two QPC potentials.

As for the SO interaction, we consider the Rashba interaction
caused by the QPC potential in Eq.\ (\ref{eq:QPC}),
that is,
\begin{equation}
H_{\rm SO} = \frac{\lambda}{\hbar} \sigma_z
\left[p_x \frac{\partial U}{\partial y} -
p_y \frac{\partial U}{\partial x} \right].
\label{eq:SO3}
\end{equation}
We choose the $z$ direction for the spin axis
(${\bm h}_{\rm SO} \parallel$ $z$ direction).

In the two-terminal geometry of Fig.\ 1(c), we consider a magnetic
field perpendicular to the $xy$ plane only in the region surrounded by
dotted line. We adopt the vector potential of $\bm{A}=(-By,0,0)$
for the orbital magnetization and neglect the Zeeman effect.

We discretize the two-dimensional space with QPC potentials and
obtain the tight-binding model. We numerically evaluate the
spin-dependent conductance, using the calculation method in
Appendix B.

We consider the following situation.
The width of quantum wires is $W=100\, \mathrm{nm}$.
The lattice constant of the tight-binding model is
$a=W/30$ (number of sites is $M=29$ in width of the
wires).
For the SO interaction, the dimensionless coupling constant
is $\tilde{\lambda}=\lambda /(2 a^2) = 0.05$,
which corresponds to $\lambda =1.171\, \mathrm{nm^2}$ in
InAs.~\cite{Winkler}
The Fermi wavelength and Fermi energy in the leads are
fixed at $\lambda_{\rm F} =W/3$ and
$\varepsilon_{\rm F}/t=2-2\cos (2\pi a/\lambda_{\rm F})
\simeq 0.382$,
respectively. (There are six conduction channels in each
lead. However, single channel is effectively coupled to the QD
owing to the QPC potential between the QD and lead.)
For the QPC potential, $L_{\rm QPC}=W_{\rm QPC} =\lambda_{\rm F}$.
$U_0=0.8 \varepsilon_{\rm F}$ at the connection to leads
S and D1,
whereas $U_0$ at the connection to lead D2
is changed from $U_0 /\varepsilon_{\rm F} =1.1$
to $0.6$ to tune the tunnel coupling $\Gamma_{\rm D2}$
in the three-terminal geometry of Fig.\ 1(b).
In Fig.\ 1(c),
the magnetic field is applied up to
$\hbar \omega_{\rm c} /\varepsilon_{\rm F} = 30 \times 10^{-4}$,
which corresponds to $B \simeq 34$ mT.

\begin{figure}
\begin{center}
\includegraphics[width=7cm]{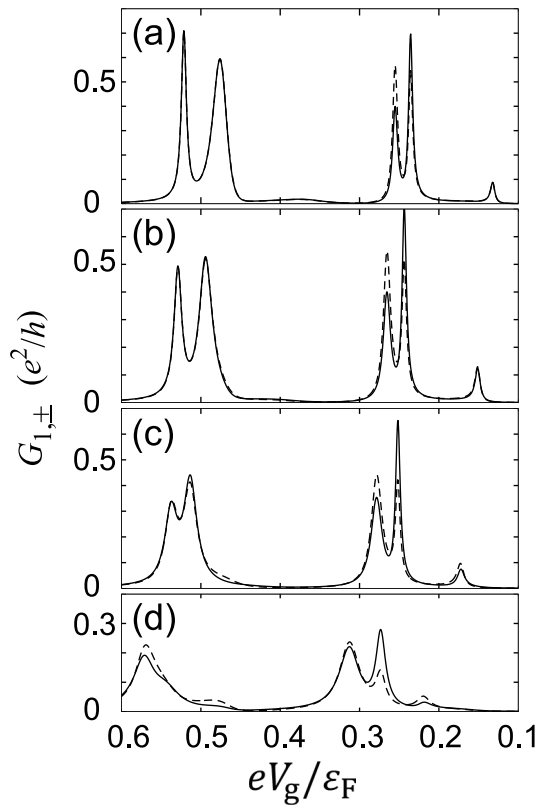}
\end{center}
\caption{
Numerical results of spin-dependent conductance
$G_{1,\pm}$ in model of Fig.\ 1(b),
as a function of electrostatic potential in the
quantum dot, $eV_{\rm g}$. No magnetic field is applied.
Solid (broken) lines indicate the conductance
$G_{1,+}$ ($G_{1,-}$) for spin $\sigma=+1$ ($-1$)
in the $z$ direction. The height of QPC potential is
$U_0=0.8\varepsilon_{\rm F}$ at the connection to
leads S and D1, whereas (a)
$U_0 /\varepsilon_{\rm F} = 1.1$,
(b) $0.9$, (c) $0.8$, and (d) $0.6$
at the connection to lead D2.
}
\label{fig:TB3cond}
\end{figure}

\subsection{NUMERICAL RESULTS}
\label{sec:TBresults}

Figure 8 presents the spin-dependent conductance
$G_{1,\sigma}$ in Fig.\ 1(b) of three-terminated QD,
in the absence of magnetic field.
$\sigma=\pm 1$ indicates the $z$-component of electron spin.
The conductance shows a peak structure as a function of
electrostatic potential in the QD, $eV_\text{g}$,
reflecting the resonant tunneling through discrete
energy levels in the QD.
Although this is similar to the Coulomb oscillation,
the peak-peak distance is underestimated  because
we neglect the electron-electron interaction.

The average of the level spacing is larger than
the level broadening in Fig.\ 8. Therefore, the
difference between $G_{1,+}$ and $G_{1,-}$ is usually
small. We observe a large spin-polarized conductance,
$G_{1,+}-G_{1,-}$, around some conductance peaks
where a few levels should be close to each other around
$\varepsilon_{\rm F}$. Look at the conductance
around $eV_\text{g}/\varepsilon_{\rm F}=0.25$.
With increasing the tunnel coupling to lead D2
by decreasing the height of QPC potential $U_0$,
the spin-polarized conductance increases, becomes
maximal, and decreases. This is in accordance with
the analytical result in section III.A although
it is hard to evaluate the level spacing and
signs of tunnel coupling around $\varepsilon_{\rm F}$.
Figure 9 plots $G_{1,+}-G_{1,-}$ as a function of
$eV_\text{g}$, which seems complicated probably due
to the interference among three levels in the QD.

\begin{figure}
\begin{center}
\includegraphics[width=6cm]{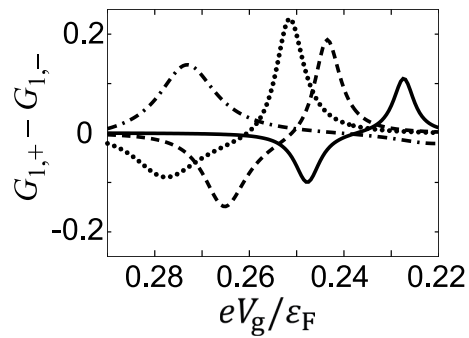}
\end{center}
\caption{
Spin-polarized conductance, $G_{1,+}-G_{1,-}$,
as a function of $eV_{\rm g}$ in model of Fig.\ 1(b).
The height of QPC potential at the connection to
lead D2 is
$U_0 /\varepsilon_{\rm F} =1.1$ (solid line),
$0.9$ (broken line),
$0.8$ (dotted line), and
$0.6$ (dotted broken line).
The other parameters are the same as in Fig.\ 8.
}
\label{fig:TB3polar}
\end{figure}

Figure 10 shows the spin-dependent conductance
$G_{1,\sigma}$ in Fig.\ 1(c) of two-terminated QD,
in the presence of magnetic field.
A large spin-polarized conductance is obtained
around $eV_\text{g}/\varepsilon_{\rm F}=-0.21$.
The difference of $G_{1,+}-G_{1,-}$ is changed
with increasing magnetic field perpendicular
to the QD.
As shown in Fig.\ 11, the absolute value of
spin-polarized conductance increases, becomes
maximal, and decreases, in accordance with
the analytical result in section III.B.

\begin{figure}
\begin{center}
\includegraphics[width=6cm]{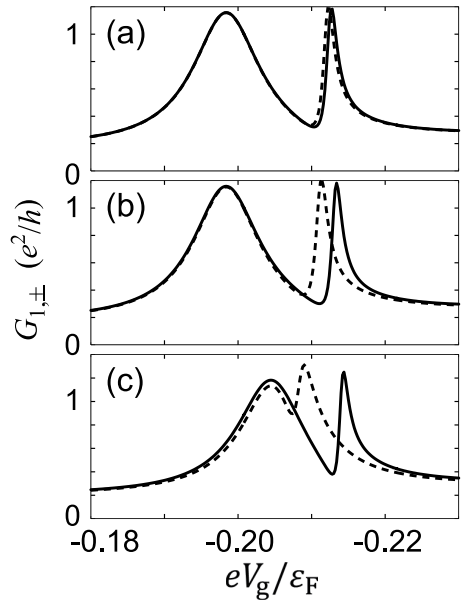}
\end{center}
\caption{
Numerical results of spin-dependent conductance
$G_{1,\pm}$ in model of Fig.\ 1(c),
as a function of electrostatic potential in the
quantum dot, $eV_{\rm g}$, in the presence of magnetic field.
Solid (broken) lines indicate the conductance
$G_{1,+}$ ($G_{1,-}$) for spin $\sigma=+1$ ($-1$)
in the $z$ direction.
The magnetic field is 
(a) $\hbar \omega_{\rm c}/\varepsilon_{\rm F}
=2 \times 10^{-4}$, (b) $10 \times 10^{-4}$,
and (c) $30 \times 10^{-4}$.
}
\label{fig:TB2cond}
\end{figure}

\section{CONCLUSIONS AND DISCUSSION}

We have studied the mechanism of SHE at a QD
with discrete energy levels in multi-terminal geometries.
We have considered a QD with SO interaction connected to $N$
external leads via tunnel barriers.
When an unpolarized current is injected to the QD from
a lead, a polarized current is ejected to others.
$N \ge 3$ ($N \ge 2$) is required in the absence (presence) of
magnetic field for the generation of spin-polarized current.

First, we have obtained the analytical expressions for the
spin-dependent conductance using a minimal model of two-level
QD. The SHE is markedly enhanced by the resonant
tunneling when the level spacing in the QD
is smaller than the level broadening due to the tunnel coupling
to the leads.
In the absence of magnetic field,
the spin polarization can be tuned by changing the tunnel coupling to
the lead other than source and drain leads in a three-terminal
geometry. A weak magnetic field can tune the spin polarization
in a two-terminal geometry.

Second, we have performed numerical studies on the tight-binding
model representing a QD with tunnel barriers to $N$ leads.
We have observed a large spin-polarized conductance at some current
peaks 
when a few energy levels in the QD are close to each other around
$\varepsilon_{\rm F}$. The numerical results are in accordance with
our analysis of the minimal model of two-level QD.

In our calculation, we have neglected the electron-electron interaction.
Therefore, our theory is applicable only around the current peaks
of the Coulomb oscillation, where the interaction is not qualitatively
important. We have also neglected the Zeeman effect.
In spite of a large g-factor in InAs
($|g| \sim 10$),~\cite{Takahashi,Kanai,Deacon,Schroer,Nadj-Perge1,Nadj-Perge2}
the Zeeman effect is smaller than the orbital magnetization
by one order of magnitude for $B \sim 40$ mT,
as estimated in Appendix A.
In the absence of SO interaction, the Zeeman effect
splits a spin-degenerate level in the QD,
which could result in the spin-polarized current by
the resonant tunneling through one of the spin-split levels.
In our situation, however, the spin splitting is much smaller
than the level broadening, and hence the spin polarization by
the Zeeman effect is negligibly small.

\begin{figure}
\begin{center}
\includegraphics[width=6cm]{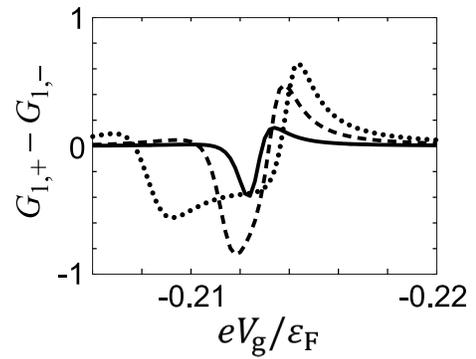}
\end{center}
\caption{
Spin-polarized conductance, $G_{1,+}-G_{1,-}$,
as a function of $eV_{\rm g}$ in model of Fig.\ 1(c)
in the presence of magnetic field.
The magnetic field is
$\hbar \omega_{\rm c}/\varepsilon_{\rm F}
=2 \times 10^{-4}$ (solid line),
$10 \times 10^{-4}$ (broken line),
and $30 \times 10^{-4}$ (dotted line).
The other parameters are the same as in Fig.\ 10.
}
\label{fig:TB2polar}
\end{figure}

We discuss a possible observation of the SHE at a QD.
Since the measurement of a spin-polarized current is usually
difficult, an alternative method is desirable.
Hamaya {\it et al}.\ fabricated InAs QDs connected
to ferromagnets.~\cite{Hamaya}
If a ferromagnet is used only for lead S and a normal metal
or semiconductor is used for the other leads,
an ``inverse SHE'' takes place.
The electric current to lead D1 is proportional to
$(1+p\cos\theta) G_{1,+}+(1-p\cos\theta) G_{1,-}$, where $p$ is
the polarization in the ferromagnet and $\theta$ is the
angle between the magnetization and ${\bm h}_\text{SO}$.
Thus $G_{1,+}$ and $G_{1,-}$ can be
evaluated by measuring the electric current with rotating the
magnetization of the ferromagnet.

The QDs are highly tunable systems. We believe that
the detailed study of the generation of spin-polarized current
at the QDs would contribute to the deeper understanding of the SHE.

\section*{ACKNOWLEDGMENT}

This work was partly supported by a Grant-in-Aid for Scientific
Research from the Japan Society for the Promotion of Science,
and by the Global COE Program ``High-Level Global Cooperation for
Leading-Edge Platform on Access Space (C12).''
T.\ Y.\ is a Research Fellow of the Japan Society for
the Promotion of Science.

\appendix
\section{Approximation in weak magnetic field}

In the presence of magnetic field and SO interaction,
the Hamiltonian in Eq.\ (\ref{eq:generalH}) is approximated to
that in Eq.\ (\ref{eq:weakH}) for the following reasons.
We consider the situation in which
the average of level spacing in the QD is $\delta \sim
\hbar^2/(m^* d^2) \simeq 1$ meV, the strength of SO interaction is
$\Delta_{\rm SO} \simeq 0.2$ meV, and magnetic field of
$\hbar \omega_{\rm c} \sim \Delta_{\rm SO}$
or smaller.
Here, $d$ is the one-dimensional size of the QD,
effective mass $m^*/m_0 \simeq 0.024$ in InAs, and
$\omega_{\rm c}=|e|B/m^*$
is the cyclotron frequency.

We choose the gauge of $\bm{A}=(\bm{B} \times \bm{r})/2$ in
the model of two-level QD. The first-order in $\bm{A}$
in Hamiltonian (\ref{eq:generalH}) gives rise to the matrix element
in Eq.\ (7). $|b| \sim \hbar \omega_{\rm c}$, as denoted in Sec.\ II.
The second order in $\bm{A}$ yields
$e^2/(8m^*) \langle i| (\bm{B} \times \bm{r})^2 |j \rangle
\sim (eBd)^2/m^* = (\hbar \omega_{\rm c})^2/\delta$,
which is smaller than the first-order term by $\hbar \omega_{\rm c}/\delta \ll 1$.

We also neglect the Zeeman effect, $H_{\rm Z} =
g \mu_{\rm B} {\bm B} \cdot {\bm \sigma/2}$,
where $\mu_{\rm B}=|e|\hbar/(2m_0)$ is
the Bohr magneton. We estimate the effect to be
$|g|\mu_{\rm B} B/2 =(|g|/4) (m^*/m_0) \hbar \omega_{\rm c}$.
Since $|g| \sim 10$ in
InAs,~\cite{Takahashi,Kanai,Deacon,Schroer,Nadj-Perge1,Nadj-Perge2}
The Zeeman term is smaller than
the orbital magnetization $|b|$ by one order of magnitude.

In the presence of magnetic field, $\bm{p}$ in $H_{\rm SO}$ is
replaced by $(\bm{p}-e\bm{A})$. In the case of Rashba interaction,
\begin{eqnarray}
H_{\rm RSO}({\bm B}) &=& \frac{\lambda}{\hbar}
\left[ (\bm{p} - e\bm{A}) \times \bm{\nabla} U \right] \nonumber \\
&=& \frac{\lambda}{\hbar} (\bm{p} \times \bm{\nabla} U)
- \frac{e\lambda}{2\hbar} \left[ (\bm{B} \times \bm{r})
\times \bm{\nabla} U \right].
\label{eq:Rashba-B}
\end{eqnarray}
The matrix element of the first term in Eq.\ (\ref{eq:Rashba-B})
is estimated to be
$(\lambda/\hbar) |\langle 2| \bm{p} \times \bm{\nabla} U |1 \rangle|
\sim (\lambda/d^2) \delta$, whereas that of the second term is
to be
$|e| \lambda/(2\hbar) |\langle i| (\bm{B} \times \bm{r})
\times \bm{\nabla} U |j \rangle|
\sim (|e| \lambda B/\hbar) \delta$. The latter is smaller than
the former by $\hbar \omega_{\rm c}/\delta \ll 1$, and thus it is
safely disregarded.
In the case of Dresselhaus interaction,
\begin{eqnarray}
H_{\rm DSO}({\bm B}) &=& \frac{\lambda'}{\hbar}
\bigl[ (\pi_y \pi_x \pi_y - \pi_z \pi_x \pi_z) \sigma_x
\nonumber \\
& & \ + (\pi_z \pi_y \pi_z - \pi_x \pi_y \pi_x) \sigma_y
\nonumber \\
& & \ + (\pi_x \pi_z \pi_x - \pi_y \pi_z \pi_y) \sigma_z \bigr],
\end{eqnarray}
where $\bm{\pi} = \bm{p} - e\bm{A}$.
The matrix element of the terms without ${\bm A}$
[Eq.\ (\ref{eq:Dresselhaus})] is estimated to be
$(\lambda^\prime \hbar^2/d^3)$ and that of the
first order in $\bm{A}$ is to be
$(\lambda^\prime \hbar/d) |e|B$. Again, the latter
is smaller than the former by $\hbar \omega_{\rm c}/\delta \ll 1$.

\section{Numerical calculation of tight-binding model}

In the model of Figs.\ 1(b) and (c), we discretize the
$xy$ plane with QPC potentials and obtain the two-dimensional
tight-binding model of square lattice.~\cite{Datta}
The lattice constant is denoted by $a$.
For the region surrounded by dotted line, the
Hamiltonian is given by
\begin{eqnarray}
H &=& t \sum_{j,l,\sigma}
\left( 4+\tilde{U}_{j,l} \right)
c_{j,l;\sigma}^\dagger c_{j,l;\sigma} \nonumber \\
&-&t \sum_{j,l,\sigma} \left(T_{j,l ; j+1,l; \sigma}
c_{j,l;\sigma}^\dagger c_{j+1,l;\sigma} +
T_{j,l ; j,l+1 ;\sigma}
c_{j,l;\sigma}^\dagger c_{j,l+1;\sigma} + {\rm h.\ c.} \right),
\label{eq:tbH}
\end{eqnarray}
where $c_{j,l;\sigma}^\dagger$ and $c_{j,l;\sigma}$
are creation and annihilation operators of an electron
at site $(j,l)$ with $z$-component of spin
$\sigma=\pm 1$, respectively. The transfer integral is
$t=\hbar^2 /(2m^* a^2)$. $\tilde{U}_{j,l}$ represents
the QPC potential and electrostatic potential in the QD,
at site $(j,l)$ in units of $t$.
The transfer term in the $x$ direction is given by
\begin{equation}
T_{j,l; j+1,l; \pm} = \left\{ 1\pm {\rm i} \tilde{\lambda}
(\tilde{U}_{j+1/2,l+1/2} -\tilde{U}_{j+1/2,l-1/2})
\right\} e^{{\rm i}2\pi \tilde{B}l},
\label{eq:xhopping}
\end{equation}
where $\tilde{\lambda} =\lambda /(2 a^2)$ is a
dimensionless strength of SO interaction and
$\tilde{U}_{j+1/2,l+1/2}$ is
the potential at the middle point between
the sites $(j,l)$ and $(j+1,l+1)$.
The magnetic field in the QD is taken into account by
the Peierls phase factor, $e^{{\rm i}2\pi \tilde{B}l}$ with
$\tilde{B}= |e|Ba^2/h$.
$\tilde{B}$ is related to the cyclotron frequency by
$\tilde{B}=\hbar \omega_{\rm c}/(4\pi t)$.
The transfer term in the $y$
direction is given by
\begin{equation}
T_{j,l ; j,l+1; \pm} =1\mp {\rm i} \tilde{\lambda}
(\tilde{U}_{j+1/2,l+1/2} -\tilde{U}_{j-1/2,l+1/2}).
\label{eq:yhopping}
\end{equation}

To randomize the discrete energy levels in the QD,
we introduce a uniformly distributed on-site energy
$w_{i,j}$ in the range of
$-W_\text{ran}/2 \leq w_{i,j} \leq W_\text{ran}/2$.
We choose $W_\text{ran}=0.2 \varepsilon_{\rm F}$.
We disregard the SO interaction induced by the random potential.

The spin-dependent conductance is numerically evaluated in the
following way. First, we define the channels in the leads 
outside of the dotted line, which
are represented by the quantum wires of width $W=(M+1)a$.
Consider a quantum wire in the $x$ direction.
There are $M$ channels, $M_{\rm cond}$ of which are
conduction modes and $(M-M_{\rm cond})$ are evanescent modes.
The wavefunction of conduction mode $\mu$ 
($\mu=1,2,\cdots,M_{\rm cond}$) is written as
\begin{eqnarray}
\psi_\mu (j,l) & = & 
\exp( {\rm i}k_{\mu}a j) u_{\mu} (l), \\
u_{\mu} (l) & = &
\sqrt{\frac{2}{M+1}} \sin
\left( \frac{\pi \mu l}{M+1} \right),
\label{eq:Wavefunc}
\end{eqnarray}
with $l=0,1,2,\cdots,M$. The wavenumber $k_{\mu}$ satisfies
$\varepsilon_\mu (k_\mu )=\varepsilon_{\rm F}$, where
the dispersion relation is given by
\begin{equation}
\varepsilon_\mu (k)= 4t - 2t\cos
\left( \frac{\pi \mu}{M+1} \right) -2t\cos (k a).
\label{eq:cosBand}
\end{equation}
The band edge, $\varepsilon_\mu (k=0)$, is located below
$\varepsilon_{\rm F}$ for the conduction modes.
The wavefunction of evanescent mode $\mu$ 
($\mu=M_{\rm cond}+1,\cdots,M$) is written as
\begin{equation}
\psi_\mu (j,l) = \exp (-\kappa_{\mu}a j) u_{\mu} (l),
\end{equation}
where $a j$ is the distance from the QD along the lead.
The band edge is located above $\varepsilon_{\rm F}$ and
$\kappa_{\mu}$ is determined from
$\varepsilon_\mu ({\rm i}\kappa_{\mu})=\varepsilon_{\rm F}$.

Next, we introduce the retarded Green function
$\hat{G}_{\sigma}(\varepsilon)$ for the inside region of
dotted line in Figs.\ 1(b) and (c).
Here, $\sigma=\pm 1$ represents the $z$ component of spin,
which is a good quantum number in Hamiltonian (\ref{eq:tbH}).
It is defined by
\begin{equation}
\hat{G}_{\sigma}(\varepsilon) = \left[
\varepsilon I -\mathcal{H}_{\sigma} - \sum_{\alpha ={\rm S,D}n}
\Sigma_\alpha \right]^{-1},
\label{eq:Green}
\end{equation}
where $\mathcal{H}_{\sigma}$ is the matrix of Hamiltonian with
spin $\sigma =\pm$. $\Sigma_\alpha$ is the self-energy due to
the tunnel coupling to lead $\alpha (={\rm S,D}n)$
and given by
\begin{equation}
\Sigma_\alpha =-t \ \tau_\alpha^\dagger U
\Lambda U^{-1} \tau_\alpha.
\end{equation}
$U=(\bm{u}_1, \bm{u}_2, \cdots, \bm{u}_M )$ is an unitary
matrix, with
$\bm{u}_{\mu}=(u_{\mu}(1), u_{\mu}(2) , \cdots , u_{\mu}(M))^{\rm T}$
in Eq.\ (\ref{eq:Wavefunc}).
$\Lambda={\rm diag}(\lambda_1,\lambda_2,\cdots,\lambda_M)$,
where $\lambda_{\mu}=\exp ({\rm i}k_{\mu} a)$ for conduction modes
and $\lambda_{\mu}=\exp (-\kappa_{\mu}a)$ for evanescent
modes. $\tau_\alpha$ is a coupling matrix between the edge of
lead $\alpha$ to the considering region;
$\tau_\alpha (l,\alpha_l)=1$ if site $\alpha_l$ is connected to
the site $l$ ($=1,2,\cdots,M$) at the end of the lead,
$\tau_\alpha (l,\alpha_l)=0$ otherwise.~\cite{Datta}

The conductance from lead S to D1 can be evaluated
separately for $\sigma=\pm 1$ of the $z$ component of spin.
The spin-dependent conductance is calculated using the formula
\begin{equation}
G_{1,\pm} =\frac{4e^2}{h} {\rm Tr} \left[ \Gamma_{\rm D1}
\hat{G}_\pm (\varepsilon_{\rm F}) \Gamma_{\rm S}
\hat{G}_\pm^\dagger (\varepsilon_{\rm F}) \right]
\label{eq:TBcond}
\end{equation}
at $T=0$, where
$\Gamma_\alpha = {\rm i}[\Sigma_\alpha -\Sigma_\alpha^\dagger ]/2$.~\cite{Datta}


\begin{thebibliography}{99}
\bibitem{com1}
In the presence of more than one conduction channel
in the leads, a spin-polarized current can be generated
in two-terminal systems without magnetic field, e.g.,
see Ref.\ \onlinecite{EKHjpsj}.
\end{thebibliography}
\end{document}